\documentclass[11pt]{arxiv-article}

\pdfoutput=1

\bibliography{References}

\usepackage{Definitions}

\usepackage{color}

\usepackage[bookmarks=false]{hyperref}

\usepackage{setspace}
\usepackage{amssymb}%
\usepackage{pifont}%

\preprint{\texttt{CERN-TH-2016-221}}

\newcommand{\OfficialTitle}{
Compensating strong coupling with large charge
}
\title{\vspace{2cm}
  {\color{Thoughtless}\Huge\textbf{\dosserif\OfficialTitle}}
}

\author{%
  \begin{minipage}{.8\linewidth}
    \vspace{1cm}
    \begin{center} \dosserif
      {\small 
        \textbf{Luis~Alvarez-Gaume}\footnote{\!\!\textsuperscript{\ding{95}}On leave of absence from the Department of Theoretical Physics, CERN.}\textsuperscript{\ding{95}},
         \textbf{Orestis Loukas}\textsuperscript{\ding{96}},
         \textbf{Domenico~Orlando}\textsuperscript{\ding{96}} and 
        \textbf{Susanne~Reffert}\textsuperscript{\ding{96}}}
    \end{center}
    \vspace{1cm}
    \authorBlock{\ding{95}}{Theory Department -- CERN,\\ 
 \textsc{ch}-1211 Geneva 23, Switzerland}
     \authorBlock{\ding{95}}{ Simons Center for Geometry and Physics,\\ State University of New York
       Stony Brook,\\ \textsc{ny}-11794-3636, \textsc{usa}}
    \authorBlock{\ding{96}}{Albert Einstein Center for Fundamental Physics\\
      Institute for Theoretical Physics\\
      University of Bern,\\
      Sidlerstrasse 5, \textsc{ch}-3012 Bern, Switzerland}
  \end{minipage}l
}

\date{} 

\begin{document}

\setstretch{1.15}

\numberwithin{equation}{section}

\begin{titlepage}

  \newgeometry{top=23.1mm,bottom=46.1mm,left=34.6mm,right=34.6mm}

  \maketitle

  \thispagestyle{empty}

  \vfill\dosserif
  
  \abstract{\normalfont \noindent
We study {some} (conformal) field theories with global
symmetries in the sector where the value of the global
charge $Q$ is large.  We find (as expected) that the low
energy excitations of this sector are described by
the general form of Goldstone's theorem in the 
non-relativistic regime.  We also derive the
unexpected result, first presented in~\cite{Hellerman:2015nra}, that
the effective field theory describing such sector
of fixed $Q$ contains effective couplings
$\lambda_{\text{eff}}\sim \lambda^b /Q^{a}$, where
$\lambda$ is the original coupling.
Hence,
large charge leads to weak coupling.  In the last section of the paper
we present an outline of how to compute anomalous dimensions {of the
$O(n)$ model} in this limit.}

\vfill

\end{titlepage}

\restoregeometry

\tableofcontents

\section{Introduction and Conclusions}

The slightly provocative title of this paper  refers to the
very intriguing results presented in~\cite{Hellerman:2015nra} where the anomalous
dimensions of operators with large global charge $J$ in certain
\acp{cft} in three-dimensions were obtained.
In most \acp{cft}, the description in terms of local Lagrangians
is not  adequate, because the theory is frequently strongly coupled,
and thus, the perturbative description
is not valid.  However, if the theory has some global symmetry
group, and if we consider it in the sector with large
values of the associated charges, the effective theory
describing those operators is found to be effectively at weak couplings.
In such a regime, quite universal results can be obtained for the anomalous
dimensions of the operators.  This paper is a first attempt to understand
the generality of these results.

\bigskip
It is well known that Goldstone's theorem presents a far
richer phenomenology when the theory is non-relativistic~\cite{Guralnik:1967zz,Nielsen:1975hm} (see also the review~\cite{Brauner:2010wm}, and~\cite{Watanabe:2013uya}). The counting of Goldstone bosons 
and their dispersion relations is more elaborate.  In fact,
if we consider relativistic field theories like \ac{qcd} 
with non-zero chemical potential for global symmetries,
(see for instance~\cite{Schafer:2001bq}) we have to consider the
theory in a non-relativistic context, and  the
low-energy excitations follow the more general form
of Goldstone's theorem. 
Relativistic theories in the sector of fixed global charges
have been also studied in the past~\cite{Schafer:2001bq, Nicolis:2011pv, Nicolis:2012vf, Nicolis:2013sga},
and it is interesting that Type I and II Goldstone bosons
appear in general\footnote{Chadha and Nielsen studied the
general non-relativistic spontaneous symmetry breaking, and
they concluded that the dispersion relation of Type-I (resp -II)
Goldstone bosons are those where $E\sim p^{2n+1}$, (resp.
$E\sim p^{2n}$), with $n$ an integer. The more common cases are 
those where $E\sim p$ and $E\sim p^2$.}.  

\bigskip
The aim of this paper is to show that when we consider quantum
field theories with global symmetries, and we study the
sector of the Hilbert space of states with large values
of the global charge, we find not only that generically
there are Goldstone excitations in the effective Lagrangian
describing that sector of the theory, but furthermore,
it seems that the effective couplings are related to the
original couplings but suppressed by powers of the global
charge.  \emph{Hence the larger the charge, the weaker the coupling
and thus the more reliable the results obtained in perturbation
theory.}

In this paper, we consider the case of the $O(N)$ vector model. Two particular results to be stressed are that
\begin{enumerate}
\item a homogeneous fixed-charge ground state (spin-$0$ operator) is only possible for a specific choice of the $O(N)$ weights\footnote{We would
    like to thank Simeon Hellerman for discussions about this point.};
\item the light spectrum of the theory around the large-charge ground state contains in addition to a single relativistic Goldstone boson also $N-1$ modes with parametrically slow propagation speeds.
\end{enumerate}

\begin{figure}
  \centering
\begin{tikzpicture}
  \draw[-latex] (-.5,0) -- (6,0) node[anchor=north west] {\(v\)};
  \draw[-latex] (0,-.5) -- (0,4) node[anchor=south east] {\(V_{\text{cl}}(v)\)};
  \draw[thick, blue] (.5,4) parabola bend (2,.5) (5.5,4) ;
  \node at (2,1) {\(\ket{v}\)};
\end{tikzpicture}
  \caption{Qualitative behavior of classical centrifugal potential, whose minimum determines the vacuum $\ket{v}$ around which we setup our perturbation theory.}
  \label{fig:barrier}
\end{figure}
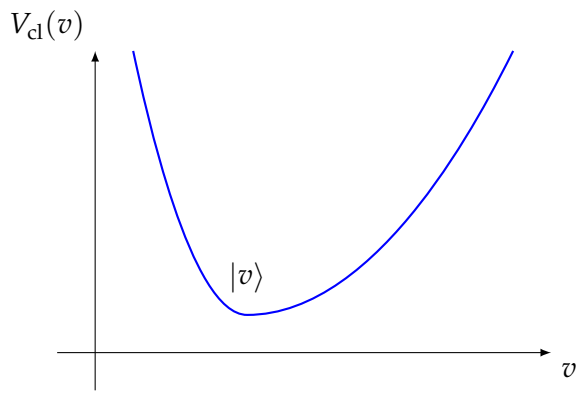

\bigskip
The organization of this paper is as follows.  Sections~\refstring{sec:o2}, and~\refstring{sec:o2n} represent a review of the general properties of the subsector
of the Hilbert space of fixed charge for a theory with a globally
conserved charge.  We will re-obtain the results that generically,
Goldstone modes are associated with the low-energy excitations
around the relevant ground states with finite charge.  Generically,
type-I and -II Goldstone fields are expected.  The ground states in fixed charge
sectors are time-dependent, hence only space-translation
invariance is assumed for the lowest energy sector.  In fact,
the effective potential in the sector of large charge is similar
to the effective potential in classical mechanics in the presence
of a central potential and fixed angular momentum.  The large
field sector is {controlled} by the original potential, but
the small field sector is suppressed by the ``centrifugal 
barrier'' provided by the large conserved charge, see
Figure~\ref{fig:barrier}.

{When} the addition of the two effects produces a 
new minimum for the field theory {there is an} associated Goldstone
excitation, which can be understood as a condensate $\ket{v}$ around
which we expand.  This seems quite a general phenomenon, {at least
  for scalar fields}.
We tailor the presentation of the rather known results in Sections~\refstring{sec:o2}, and~\refstring{sec:o2n} 
in a way adapted to the computations presented in the following
sections, in particular the study of the dispersion relations
associated to the low-energy excitations.  We elaborate on both Abelian and non-Abelian symmetries in scalar theories, and explain their
differences.  In Section~\refstring{sec:supp}, we show how the restriction
of the theory to states of large global charges involves the
study of effective Lagrangians where the originally finite (or even large) coupling constants
are suppressed by  values of the large global charge:
$\lambda_{\rm eff}\sim \lambda^b/Q^{a}$, where $\lambda$ is
the original coupling, $Q$ is the value of the global charge,
and $a$, \(b\) are positive exponents.  Finally in Section~\refstring{sec:confdim}, we
go back {and generalize to any $O(N)$ model} the computation of anomalous dimensions presented
in~\cite{Hellerman:2015nra}, and provide support for the arguments and conclusions
in that paper. 

\bigskip
It is remarkable that the formulae obtained for the anomalous
dimensions of charged operators agree extremely well with
their numerical (non-perturbative, in principle) values obtained
in~\cite{Hasenbusch} (see also~\cite{Kos:2015mba,Nakayama:2016jhq}), even for small values of the
global charge (see Section~\refstring{sec:confdim}).  This seems to imply that
the analytic expressions in~\cite{Hellerman:2015nra} for anomalous dimensions 
are such that the terms proportional to positive powers
of $J^{1/2}$ are universal.  From that, it becomes clear that much
remains to be understood in the study of quantum
field theories in their large global charge sectors.
We plan to come back to many of the open questions left open, like the universality of the results in~\cite{Hellerman:2015nra,Hasenbusch}, and what happens when we include other fields apart from scalars in the near future.

\section{Systems with Abelian global symmetry at fixed charge}\label{sec:o2}

In this section, we are concerned with a very general system exhibiting a conserved Abelian global charge. 
First, we discuss the implications of fixing the charge in the classical case and then in the quantum version.
Using first principles it is shown that the existence of a Goldstone boson always follows from charge fixation.

\subsection{Classical analysis}

We begin by studying a general classical system described by Hamiltonian $H$ with a conserved Abelian global symmetry:
\begin{equation}
  \pb{H}{Q}=0\,.
\end{equation}
In order to fix the charge, we impose the constraint\footnote{Note that we work at finite volume.}
\begin{equation}\label{eq:fixQ}
Q=\int \rho\, \dd x=\overline Q =\mathrm{const}\,.
\end{equation}
This is a first-class constraint and generates the gauge transformation
\begin{equation}
  \delta_\epsilon f= \pb{f}{\epsilon Q}\,,
\end{equation}
where $f$ is a function in phase space. Clearly, $\delta_\epsilon$ leaves the Hamiltonian invariant.
The zero-mode contribution to $Q$ is $\rho$ and $\chi$ is its canonical conjugate,
\begin{equation}
  \pb{\chi}{Q} = 1\,,
\end{equation}
so that
\begin{equation}
  \delta_\epsilon\chi = \epsilon\,,
\end{equation}
while all other variables are gauge invariant.\footnote{This is not necessary but results in a simplification.} We now have the phase space coordinates $(p_i,q_i)$, $(\rho, \chi)$. They fulfill the usual Hamilton's equations
\begin{align}
  \dot p_i &= \pb{p_i}{H},& \dot q_i &=\pb{q_i}{H},\\
  \dot \chi &=\pb{\chi}{ H},& \dot\rho &=\pb{\rho}{H}=0,
\end{align}
plus the constraint Eq.~(\ref{eq:fixQ}). For concreteness, let us
consider a Hamiltonian that is quadratic in the momenta and the
gradient of the positions\footnote{This is called a natural Hamiltonian
  system.}:
\begin{equation}
  H=\tfrac{1}{2}\sum_{k=0}^Nf_k(q)p^2_k + \tfrac{1}{2}\sum_{k=0}^Ng_k(q)(\nabla q_k)^2 + V(q) ,
\end{equation}
with $p_0=\rho$, $q_0=\chi$ and $f_k,\ g_k$ functions.
We want to find the ground state of this system. As the Hamiltonian is the sum of positive terms, we need to set them each to zero separately. Because of the constraint, $\rho\neq0$, but we are free to set
\begin{align}
\nabla q_i &=0, & \nabla \chi &= 0,& p_i &=0, \hspace{2em} i =1,\dots,N\,.
\end{align}
Since nothing depends on the position anymore, the constraint Eq.~(\ref{eq:fixQ}) becomes
\begin{equation}\label{eq:volQ}
  \int \rho \, \dd{x} = \text{vol.} \times \bar \rho =\overline Q\,.
\end{equation}
For the rest of this paragraph, we use $\rho=\bar \rho$. The remaining \ac{eom} are
\begin{align}
\dot p_i&=\del_i f_0\,\bar\rho^2+\del_iV=0 \,,\\ 
\dot q_i&=0 \,,\\
\label{eq:dotchi}\dot \chi&=f_0(q_i)\bar\rho\,.
\end{align}
They are solved by 
\begin{align}\label{eq:classSol}
  p_i &=0\,, & q_i &=\bar q_i(\bar\rho)\,, & \chi &= f_0(\bar q_i(\bar \rho))\bar\rho t = \mu(\bar\rho)t\,,
\end{align}
where $\bar q_i$ and $\mu(\bar\rho)$ are constants. Note that we used the gauge freedom to set $\chi(0)=0$. This solves the classical problem. %

\subsection{Quantization via variational approach}

In the following, we want to quantize the above classical system using a variational approach.\footnote{A perturbative approach where perturbations around the classical ground state are quantized leads to the same result.} We want to find a state $v$ that minimizes
\begin{equation}
  \expval{H}{v}
\end{equation}
under the constraints
\begin{align}
  \braket{v} = 1 && \text{and} && \expval{\rho}{v} = \bar\rho \,.
\end{align}
We introduce the Lagrange multipliers $E$, $m$ and minimize
\begin{equation}
  \expval{ H - E_0 - m \rho}{v}.
\end{equation}
The solution is 
\begin{equation}
  \pqty{H - E_0 - m \rho} \ket{v} = 0 \,.
\end{equation}
In order to reproduce the classical solution Eq.~(\ref{eq:classSol}), we must have
\begin{equation}
  \expval{\dot \chi}{v} = \mu \,,
\end{equation}
where $\mu$ is the value found in Eq.~\eqref{eq:classSol}. Now,
\begin{equation}
  \expval{\dot \chi}{v} = \expval{\comm{\chi}{H}}{v } = m \expval{\comm{\chi}{\rho}}{v},
\end{equation}
and since $\chi,\,\rho$ are canonically conjugate, we obtain 
\begin{equation}
  m=\mu\,.
\end{equation}
At the end of the day we find that the quantum Hamiltonian is given by
\begin{equation}
  \mathcal{H} = H - \mu \rho - E_0 \,,
\end{equation}
where $\mu$ is now fixed and not a Lagrangian multiplier anymore, acting as a fixed chemical potential. This reproduces the situation discussed in~\cite{Nicolis:2011pv} and automatically assures the existence of a Goldstone boson, as we will summarize in the next section.

\subsection{Existence of Goldstone modes}

Based on the earlier results of~\cite{Guralnik:1967zz,Nielsen:1975hm}, a generalization of the standard relativistic Goldstone's theorem was discussed in~\cite{Nicolis:2011pv,Nicolis:2012vf,Nicolis:2013sga}. Since this version of the theorem is crucial for a deeper understanding of our results, we review its basics here.

Assume that we start from a relativistic theory with Hamiltonian $H$ and conserved charge $Q$, \emph{i.e.} $\comm{H}{Q}=0$\,. The ground state $\ket{v}$ is taken to break both this symmetry as well as time-translation invariance, but in a controlled manner, such that the combination
\begin{equation}
  \label{eq:ymmetryBreakingVacuum}
  \left( H- \mu  Q \right) \ket{v} = 0
~
\end{equation}
is invariant.
As $H$ is explicitly time-independent in a relativistic theory on $\setR_t \times \setR^{d}$, $\mu$ has to be  constant. 
An example of such state is the one with finite charge density,
$\expval{\rho}{v} = \rho_0$\,, considered in the previous paragraph.
From the Lorentz algebra one can then show that all Lorentz boosts are broken by such vacuum choice, while spatial invariance remains unaffected.

Under those assumptions we can  prove the existence of a Goldstone boson.
This proceeds along similar lines as the familiar relativistic case.
Consider for a local operator $A(x)$ the expectation value
$\expval{\comm{Q}{A(0)}}{v}$, which is non-vanishing due to symmetry breaking.
Charge conservation implies that this matrix element
is a non-vanishing constant:
\begin{equation}
  \int \dd[d-1]x \expval{e^{i( P\cdot X + H t)}  \rho \ e^{-i( P\cdot X+ H t)} A(0)}{v} - \text{h.c.} =  \text{const} \neq 0 
\, ,
\end{equation}
where $X,P$ are the position and momentum operators.\footnote{Recall that $H$ does not annihilate the vacuum while $P$ does because of the spatial homogeneity of the ground state. For this reason, similar arguments apply here as in the standard relativistic case, allowing us to safely disregard purely spatial surface terms~\cite{Nicolis:2011pv}.}
Using \eqref{eq:ymmetryBreakingVacuum} and developing on a complete set of simultaneous  $P$ and $(H - \mu Q)$ eigenstates we find
\begin{equation}
  \text{const} = \sum_p\delta^{(d-1)}(p) \matrixelement{v}{\rho\, \text{e}^{-i( H - \mu Q)t}}{p} \matrixelement{p}{A(0)}{v} - \text{h.c.}
\,,
\end{equation}
Now, if at $p\rightarrow 0$ only the state $\ket{ p} = \ket{v} $ existed in the spectrum, the matrix element would vanish because both $\rho$ and $A$ are Hermitian operators.
Therefore, we conclude that in order for our matrix element to be time-independent and non-vanishing, there should be another state $ \ket{\chi(p)} $ with the property
\begin{equation}
  \lim_{p \to 0} \pqty{H - \mu Q} \ket{\chi(p)} = 0
\,.
\end{equation}
This state corresponds to the Goldstone boson.

In fact, the authors in~\cite{Nicolis:2011pv} argue that the leading term of this Goldstone field will be 
\begin{equation}
\chi \sim \text{const} \times t
\,,
\end{equation}
at least as long as the charge density $\rho_0$ is assumed to be small.
However, as we show in this paper, this form for the Goldstone fluctuations is true especially when the charge density is large.
Also in the same work, it is correctly observed that 
as time is singled out, we should generically expect a non-trivial dispersion relation, $\omega_\chi(p) \neq p$, for the leading Goldstone field.
In our setup, this is calculated in Eq.~\eqref{O2:QuadraticHamiltonian_Diagonal}.

\section{The $O(2n)$ vector model at fixed charge}\label{sec:o2n}

Instead of only focusing  on the Abelian case of the $O(2)$ model as in~\cite{Hellerman:2015nra}, we will discuss here the general case of the $O(2n)$ vector model where we can fix up to $n$ charges of the global symmetry\footnote{The case \(O(2n+1)\) is completely analogous.}. As we will deduce in the following, despite the existence of $k$ fixed charges, a single parameter $\mu$ acts as a chemical potential, just as in the Abelian case discussed in Section~\ref{sec:o2}. We will also see that the $O(2)$ sector leads to a relativistic Goldstone boson as before, while the remaining $k-1$ fixed charges of the non-Abelian sector give rise to non-relativistic Goldstone bosons with effective mass $\mu$, all other modes being massive. Finally, we show that in the limit of large charge, all interaction terms are suppressed by $\mu \gg 1$. As an application of our formalism, we calculate the conformal dimension of the $O(2n)$ model in three dimensions, extending the result found in~\cite{Hellerman:2015nra}.

\subsection{Classical analysis}
Let us consider the Lagrangian of the  $O(2n)$ vector model (summation over repeated indices implied),
\begin{align}
\label{eq:O2n_LagrangianDensity}
  \mathcal{L} &= \tfrac{1}{2}\del_\mu\phi^a\del^\mu\phi^a-\tfrac{1}{2}V(\phi^a\phi^a), & a&=1,\dots,2n
\,,
\end{align}
in %
$\mathbb R_t\times\mathbb R^{d-1}$.
We want to fix $k\leq n$ of the charges and study the resulting effective action. First we look at the classical problem. Using the fact that
\begin{equation}
U(n)\subset O(2n),
\end{equation}
we introduce complex variables
\begin{align}
  \varphi_1 &= \frac{1}{\sqrt 2} \left(\phi_1 + i \phi_ 2 \right)\,, &  \varphi_2 &= \frac{1}{\sqrt 2} \left( \phi_3 + i \phi_4\right)\,, & \dots,
\end{align}
so that the $k$ $U(1)$ generators act as rotations:
\begin{equation}
  \pb{\varphi_i}{\epsilon_j Q_j} = \epsilon_j\delta_{ij} \varphi_i \,.
\end{equation}
Like in the Abelian case (Eq.~(\ref{eq:volQ})), we impose the conditions
\begin{equation}
  \int \dd[d-1]{ x} \rho_i= \overline Q_i = \text{vol.} \times \bar \rho_i \,,
\end{equation}
where the $\bar \rho_i$ are fixed. By the argument above, we find that the homogeneous solution, which corresponds to choosing a vector in the maximal torus, is given by
\begin{equation}
  \begin{cases}
\varphi_i = \tfrac{1}{\sqrt2} A_i\,e^{i\mu t}, & i=1,\dots,k \,,\\
\varphi_{k+j} = 0, & j=1,\dots,n-k \,,
\end{cases}
\end{equation}
where $A_i$ and $\mu$ depend on the fixed charges $\bar\rho_i$:
\begin{align}
\bar\rho_i &= A_i^2\sqrt{V'(A_1^2+\dots+A_k^2)} \,,\\
\mu &= \sqrt{V'(A_1^2+\dots+A_k^2)} \,.
\label{ChargeDensity_ChemicalPotential_Relations}
\end{align}
$\mu$ is again the equivalent of the one found in Eq.~\eqref{eq:classSol} in the classical Abelian context. The key observation is that for a homogeneous solution, the phase $\mu$ is the same for all fields, even if all the charges $\bar\rho_i$ are different (but not vanishing).
Using the variational approach of Sec.~\ref{sec:o2}, we find that the corresponding quantum problem is the diagonalization of 
\begin{equation}\label{eq:Hmu}
H-\mu(\rho_1+\rho_2+\dots+\rho_k) \,,
\end{equation}
where $\mu$ plays the role of a fixed chemical potential.
For later convenience we define
\begin{equation}
  v^2 = \sum_{i=1}^k A_i^2  = \frac{1}{\mu} \sum_{i=1}^k \bar \rho_i = \frac{\bar \rho}{\mu} \,. 
\end{equation}

\medskip
Note that both \(v\) and \(\mu\) are increasing functions of the charge \(\rho\).
Assuming that \(\rho\) is the dominant scale\footnote{This is
  automatic at the conformal point, but we will in any case assume
  that possible scales in \(V\) are much smaller than the one fixed by
  \(\bar \rho\).
}, using the fact
  that dimensionally, \([\rho] = d - 1\), \([\mu] = 1\) and \([v] = d/2
  - 1\), we can write
  \begin{align}\label{eq:whymu}
  \mu = \order{\rho^{1/(d-1)}} && \text{and} && v =
  \order{\rho^{(d-2)/(2(d-1))}} \,.
\end{align}
  This means that
  for \(d>2\), there is no problem. This is consistent with the Coleman--Mermin--Wagner theorem (no
  spontaneous breaking for \(d=2\)).
\subsection{Symmetries and counting of the Goldstone modes}
\label{sec:symm-count-goldst}

To study the symmetries of the problem, we start from the Hamiltonian in Eq.~(\ref{eq:Hmu}) and pass to the Lagrangian formalism, resulting in
\begin{multline}
\label{O2n_Lagrangian}
  \mathcal{L}_\mu = \sum_{k=1}^k (\del_t - i\mu) \varphi_i^* (\del_t+i\mu)\varphi_i + \sum_{i=k+1}^n \dot\varphi^*_i\dot\varphi_i \\
- \sum_{k=1}^n \nabla \varphi_i^*\nabla\varphi_i - V(2|\varphi_1|^2+\dots+2|\varphi_n|^2) \,.
\end{multline}
The $\mu$-dependent term is 
\begin{equation}
i\mu \sum_{i=1}^k \left(\dot\varphi_i^*\varphi_i-\varphi_i^*\dot\varphi_i \right)= i\mu(\dot{\vec \varphi}^\dagger \cdot \vec \varphi - {\vec \varphi}^\dagger\cdot \dot{\vec \varphi} ) \,,
\end{equation}
where $\vec{\varphi}=(\varphi_1 \dots \varphi_k)$  is invariant
under $\vec{\varphi} \mapsto U \vec{\varphi}$ if $U^\dagger U=\Id$. The remaining $(2n-2k)$ fields are spectators.
Since $\mu_i=\mu$ $\forall i$, independently of the $\bar\rho_i$, the
system preserves $O(2n-2k)\times U(k)$ symmetry.\footnote{The group \(O(2k)\) preserves only the combination 
\(\sum_{i=1}^{2k}  \phi_{i}' \phi_{i}   \) while \(U(k)\) preserves both
\(\sum_{i=1}^{2k}  \phi_{i}' \phi_{i}   \) and \(\sum_{i=1}^k  \left(\phi_{2i - 1}' \phi_i - \phi_{2i-i}  \phi_i'\right)\), which is the new term appearing in the Lagrangian.}

We know from the classical analysis that the vacuum corresponds to
\begin{align}
\begin{cases}
  \expval{\varphi_i} = \tfrac{1}{\sqrt 2}A_i, & i=1,\dots,k \,,\\
  \expval{\varphi_i} = 0, & i=k+1,\dots, n \,.
\end{cases}
\end{align}
This vacuum spontaneously breaks $U(k)$ to $U(k-1)$.
To see this, note that we can
rotate the vector $\expval{\vec\varphi} = \tfrac{1}{\sqrt 2} (A_1, \dots, A_k, 0, \dots )$ into
\begin{equation}
(M \oplus \Id_{N-k}) 
\expval{\vec\varphi} = (0, \dots, 0, \sqrt{\tfrac{A_1^2+\dots+A_k^2}{ 2}},0, \dots) =  (0,
\dots, 0, \tfrac{v}{\sqrt 2} ,0, \dots) \,,
\end{equation} 
where $M\in U(k)$ is a constant matrix that depends on the $A_i$. It is now clear that $\vec\varphi_0$ is invariant under transformations
\begin{equation}
M^{-1}\begin{pmatrix}\tilde U & 0\\
0 & 1 \end{pmatrix}M \,,
\end{equation}
where $\tilde U\in U(k-1)$.
We have now found the breaking pattern 
\begin{equation}
O(2n-2k)\times U(k) \to O(2n-2k)\times U(k-1)
\end{equation}
and we can compute the dimension of the coset
\begin{equation}
\dim G/H = \dim U(k) - \dim U(k-1) = k^2 - (k-1)^2 = 2k-1 \,.
\end{equation}
In a relativistic system, this would be the end of the story but by
fixing the charge, we are breaking Lorentz invariance, which leads in
general to fewer Goldstone bosons~\cite{Nielsen:1975hm,Watanabe:2013uya}.

\subsection{Semi-classical analysis and dispersion relations}

In order to count the Goldstone bosons and to study their properties, we
can start with a semiclassical analysis.
It is convenient to use the matrix \(M\) above to rotate the ground
state and expand around 
\begin{equation}
  M \expval{\vec \varphi} = \big(  \underbrace{0\,,\, \dots\,,\, 0}_{k-1}\,,\, \tfrac{v}{\sqrt 2} \,,\, \underbrace{0\,,\, \dots\,,\, 0}_{n-k} \big)
\,. 
\end{equation}
Here, we distinguish two interesting sectors. %
The first \(k - 1\) fields are expanded %
around \(\varphi_i = 0\), while the $k$-th is expanded around
\(\varphi_k = \frac{v}{\sqrt 2}\).

In this latter sector (which we will refer to as
the \(O(2)\) sector) we parameterize the fluctuations as
\begin{equation}
\label{O2_FieldParametrization}
  \varphi_k = \tfrac{1}{\sqrt 2}\,e^{i\mu t +i\hat\phi_{2k}/v}  \pqty{v+ \hat\phi_{2k-1}}
  \,,
\end{equation}
where $\hat\phi_{2k-1}\,,\,\hat\phi_{2k}$ are real-valued field operators.
The \(O(2)\) symmetry which is spontaneously broken by the \ac{vev} is realized as a
linear shift for \(\hat \phi_{2k}\):
\begin{equation}
  \begin{cases}
  \hat \phi_{2k-1} \to \hat \phi_{2k-1}\\
  \hat \phi_{2k} \to \hat \phi_{2k} + \theta
  \,,
\end{cases}
\end{equation}
which implies that an $O(2)$-invariant potential cannot depend on
$\hat\phi_{2k}$.

\bigskip

For the other fields \(\varphi_i\), \(i = 1, \dots, k-1\), forming the
\(U(k-1)\) sector, we choose instead
\begin{equation}
\label{NonAbelian_FieldParametrization}
  \varphi_i = e^{i \mu t} \hat \varphi_i \,,  
\end{equation}
where $\hat \varphi_i $ denotes complex-valued field operators.
The (unbroken) \(U(k-1)\) symmetry is then realized as
\begin{equation}
  \hat \varphi_i \mapsto \tilde U_i^{\phantom{i} j}  \hat \varphi_j \,.
\end{equation}
The two parameterizations agree for large \(v\), since 
\begin{equation}
e^{i\mu t
  +i\hat\phi_{2k}/v} (v+ \hat\phi_{2k-1}) = e^{i \mu t} (v + \hat
\phi_{2k - 1} + i \hat \phi_{2k}) + \order{1/v} \,.
\end{equation}
This is however true only up to quadratic terms, when discussing interactions, they lead to different results.

Now, we can rewrite the Lagrangian density \eqref{O2n_Lagrangian} using the parameterizations %
in Eq.~\eqref{O2_FieldParametrization} and Eq.~\eqref{NonAbelian_FieldParametrization}\,:
\begin{multline}
  \mathcal{L} = \sum_{i=1}^{k-1} ( \del_t - i \mu ) \varphi_i^* (\del_t + i \mu) \varphi_i  +  \frac{1}{2} \dot \phi_{2k-1} \dot \phi_{2k-1} + \sum_{i=k+1}^n \dot \varphi_i^* \dot \varphi_i \\
  + \frac{1}{2}\pqty{v + \phi_{2k-1}}^2 \pqty{\pqty{\mu + \frac{\dot \phi_{2k}}{v}}^2 - \frac{(\nabla \phi_{2k})^2}{v^2}  }   
  -  \sum_{i=1}^{n-1} \nabla \varphi_i^* \nabla\varphi_i - \frac{1}{2} (\nabla \phi_{2k-1})^2\\
  - 
  \tfrac12 V \left( 2\abs{\varphi_1}^2 + \dots + 2\abs{\varphi_{k-1}}^2 +  \abs{ v + \phi_{2k-1} }^2 + 2\abs{\varphi_{k+1}}^2 + \dots + 2\abs{\varphi_{n}}^2 \right) \,,
\end{multline}
where we have dropped the hat for ease of notation.\footnote{It can be
convenient to think of this action as resulting from the Kaluza--Klein
reduction of a plane wave geometry for particles with momentum \(\mu\) in the
extra dimension. More precisely, we can write 
\(\mathcal{L} = \partial_M \Phi_i^* G^{MN} \partial_M \Phi_i + V(\abs{\Phi}^2)\),
where \(G_{MN} \dd{x^M} \dd{x^N} = 2\dd{t}\dd{y} + \dd{y}^2 +
\dd{\vec{x}}^2\) and \(\Phi_i = \phi_i(x^\mu) e^{i \mu y} \).}
Developing at second order in the fields around the vacuum we find:
\begin{multline}
  \mathcal{L}^{(2)} =  \sum_{i=1}^k (\del_t - i \mu)\varphi_i^* (\del_t + i \mu)\varphi_i +\sum_{i=k+1}^n \dot \varphi_i^\ast \dot \varphi_i - \sum_{i=1}^n \nabla \varphi_i^* \nabla \varphi_i
  \\ 
  - \sum_{i = 1}^n \mu^2 \varphi_i^* \varphi_i -  \frac{2c^2}{1-c^2} \mu^2  \phi_{2k-1}^2 \,,
\end{multline}
where we used the fact that \(\mu^2 = V'(v^2)\) (relation \eqref{ChargeDensity_ChemicalPotential_Relations}) 
and for later convenience
we have introduced the dimensionless parameter \(c\) to  rewrite
\(V''(v^2)\) as\footnote{For \(V(\varphi) = \abs{\varphi}^4\), we have
  \(c = 1/\sqrt{3}\). For \(V(\varphi) = \abs{\varphi}^6\) we have
  \(c= 1/\sqrt{2}\).}
\begin{equation}
  V''(v^2) = \frac{2c^2}{1-c^2}\frac{\mu^2}{v^2}
  \,.
\end{equation}
Note that $V''(v^2)>0$ implies $c<1$.
It is clear that the fields \(\varphi_i\), \(i = k+1, \dots, N\) are a
collection of \(N - k\) massive complex scalars with mass \(\mu\), so
from now on we will concentrate on the other \(k\) complex scalars.

As usual, we pass to Fourier space and define the inverse propagator \(\Delta^{-1}(p)\) from the quadratic part of the action, namely
\begin{equation}
\int \dd[d]x\, \mathcal{L}^{(2)} = %
 \int \frac{\dd[d]p}{(2\pi)^{d} }\,
 ( \varphi_1^*(-p) \dots \varphi_k^*(-p))  \Delta^{-1}(p) \begin{pmatrix} \varphi_1(p) \\ \vdots \\ \varphi_k(p) \end{pmatrix}
 \,.
\end{equation}
One recognizes that \(\Delta^{-1}(p)\) is a block-diagonal matrix. For each
of the first \(k-1\) complex scalars $\varphi_i$\,, we have a \(2 \times 2 \) block
\begin{equation}
  \Delta^{-1}_i(p) = \begin{pmatrix}
    \tfrac12\left(\omega^2  - p^2 \right) & i \omega \mu \\
    -i \omega \mu & \tfrac12\left(\omega^2  - p^2 \right)
  \end{pmatrix},
\end{equation}
while the $k$-th field is different because of the mass term for its real component
\(\phi_{2k -1}\)\,:
\begin{equation}
  \Delta_k^{-1}(p) = \begin{pmatrix}
    \omega^2  - p^2 - \frac{4 c^2 \mu^2}{1-c^2} & 2 i \omega \mu \\
    -2 i \omega \mu & \omega^2 - p^2
  \end{pmatrix} \,.
\end{equation}
The determinant of the inverse propagator for $\phi_i$, $i=1,...,k$ is
\begin{multline}
  \det(\Delta^{-1}(p)) 
  = \prod_{i=1}^{k}   \det(\Delta^{-1}_i(p))    =
  \frac{1}{16 \pqty{1 - c^2}} \pqty{ \tfrac14\pqty{\omega^2 - p^2}^2 - \omega^2 \mu^2}^{k-1} \\
   \times \pqty{\pqty{1-c^2} \pqty{\omega^2 -p^2}^2 - 4 \mu^2 \pqty{\omega^2 - c^2 p^2 }}
  \,.
\end{multline}
The dispersion relations of the quasi-particle eigenstates are obtained as the roots of the
equation \(\det(\Delta^{-1}(p)) = 0\)\,:
\begin{align}
  \omega &= \sqrt{p^2 + \mu^2} \pm \mu & \text{\(k-1\) times}\\
  \omega_\pm &= \sqrt{p^2 + \frac{2 \mu}{1-c^2}\pqty{\mu \pm \sqrt{\pqty{1-c^2}^2 p^2 + \mu^2}}}
\end{align}
As we have seen in Eq.~\eqref{eq:whymu}, $\mu$ is large for large $\rho$ and is the most convenient expansion parameter. Expanding for large \(\mu\) we find:
\begin{align}
   \omega^2 &= \left( -\mu + \sqrt{p^2 + \mu^2} \right)^2 = \frac{p^4}{4\mu^2} - \frac{p^6}{8\mu^4} + \order{\mu^{-6}} & \text{$k-1$ times}\\
   \omega^2 &= \left( \mu + \sqrt{p^2 + \mu^2} \right)^2 = {4\mu^2} + 2p^2 + \order{\mu^{-2}} & \text{$k-1$ times}\\
  \omega_-^2 &= c^2 p^2 + \frac{\pqty{1-c^2}^3 p^4}{4 \mu^2} + \order{\mu^{-4}}   & \text{one time}\\
  \omega_+^2 &= \frac{4 \mu^2}{1-c^2} + \pqty{2 - c^2} p^2 + \order{\mu^{-2}}  &\text{one time.}
\end{align}

Even if Lorentz invariance is broken in the sector of fixed charge, the overall theory remains Lorentz invariant, which is reflected in the fact that $c<1$.

To summarize, using the notation of~\cite{Nielsen:1975hm}, we find that fixing $k$ out of $n$ charges leads to
\begin{itemize}
\item one relativistic Goldstone boson with speed of light \(c < 1\)\,,
\item $k-1$ non-relativistic Goldstones with mass $\mu$ and dispersion $\omega=\frac{p^2}{2\mu}+\dots$\,,
\item one massive state with mass $\frac{2 \mu}{\sqrt{1-c^2}}$\,,
\item $k-1$ massive states with mass $2\mu$\,,
\item $2n-2k$ massive states with mass $\mu$\,.
\end{itemize}
In condensed matter language, the system has one phonon and $k-1$
magnons.

\medskip
Now we can come back to the results of the previous
Section~\ref{sec:symm-count-goldst}. We found that the \(U(k)\)
symmetry is spontaneously broken to \(U(k-1)\), so that the coset has
dimension \(\dim(G/H) = 2 k - 1\).
Now we know that there is one relativistic Goldstone and \(k - 1\)
non-relativistic ones. In the language of~\cite{Nielsen:1975hm}, they
are of type I and II respectively. Goldstones of type II count double,
and in fact:
\begin{equation}
  1 + 2\times (k-1) = 2 k - 1 = \dim(G/H) \,.
\end{equation}

\subsection{Canonical quantization of the non-Abelian sector}

Until here, we have used semi-classical arguments to discuss the existence and the counting of the Goldstone modes. In the following, we will present a completely quantum description starting from first principles, using canonical quantization. We are able to diagonalize the resulting quantum Hamiltonian and read off the Goldstone modes from there. 

We treat the Abelian and non-Abelian sectors separately, because of the choice of \ac{vev}. For technical reasons, the non-Abelian sector is simpler, so we start out with the non-Abelian case.

\medskip
The quadratic Hamiltonian in the $\varphi_i$, \(i =1, \dots, k - 1\) is given by
\begin{equation}
  \mathcal{H}^{(2)}_i = \pi_i^* \pi_i + \nabla \varphi_i^* \nabla \varphi_i + \mu^2 \varphi_i^* \varphi_i - \mu ( \pi_i \varphi_i - \pi_i^* \varphi_i^* ) \,.
\end{equation}
In order to diagonalize it, we go to Fourier space and expand in terms
of canonical operators:
\begin{align}
  \varphi_i (p) &= \frac{1}{\sqrt{2\tilde\omega(p)}}(a_i(p) + b^{\dagger}_i(-p)) \,,\\
  \pi_i(p) &= -i \sqrt{\frac{\tilde\omega(p)}{2}}(a_i(p) - b^{\dagger}_i(-p)) \,.
\end{align}
The Hamiltonian becomes
\begin{multline}
  \mathcal{H}^{(2)}(p) = \pqty{\tilde\omega(p) - \mu} a_i^\dagger (p) a_i(p) + \pqty{\tilde\omega(p) + \mu} b_i^\dagger(p) b_i(p) + \\
+ \left( -\tilde\omega(p) + \frac{p^2+\mu^2}{\tilde\omega(p)} \right) \pqty{a_i(p)b_i(-p) + a_i^\dagger(p)b_i^\dagger(-p)} \,.
\end{multline}
and it is diagonal if $\tilde\omega^2= p^2+\mu^2$:
\begin{equation}
  \mathcal{H}^{(2)}_i(p)=\pqty{\sqrt{p^2+\mu^2}-\mu} a_i^\dagger (p) a_i(p) + \pqty{\sqrt{p^2+\mu^2}+\mu} b_i^\dagger(p) b_i(p) \,.
\end{equation}
We have broken Lorentz invariance, and with it the symmetry between
particles and antiparticles. For $\mu \gg 1$, $a$ is a
non-relativistic Goldstone with $\omega^2\sim\frac{p^2}{2\mu}$ and $b$
is massive. For later convenience we write once more the explicit
expression for the fields in terms of the oscillators and \(\mu\):
\begin{equation}\label{eq:varphi}
  \varphi_i (p) = \frac{1}{\sqrt{2} \pqty{p^2 + \mu^2}^{1/4}}  \pqty{a_i(p) + b^{\dagger}_i(-p)} \sim \frac{1}{\sqrt{2 \mu}}  \pqty{a_i(p) + b^{\dagger}_i(-p)}
  \,.
\end{equation}
  
\bigskip

Another way of looking at the problem is to write the Lagrangian
\begin{equation}
  \mathcal{L}^{(2)}_i =  \pqty{\del_t-i\mu} \varphi_i^* \pqty{\del_t + i \mu} \varphi_i -\mu^2 \varphi_i^* \varphi_i - \nabla\varphi_i^*\nabla\varphi_i \,.
\end{equation}
If $\mu\gg\del_t$, the Lagrangian becomes the one of the massless Schrödinger particle:
\begin{equation}
  \mathcal{L}^{(2)}_i = i \mu \pqty{\dot\varphi_i^* \varphi_i - \varphi_i^* \dot\varphi_i} - \nabla \varphi_i^* \nabla\varphi_i \,,
\end{equation}
which has the same dispersion relation we found for the Goldstone. The
term $\mu(\rho_1 + \dots + \rho_k)$ acts like a Berry's phase and when it
dominates, we get only one classical Goldstone particle instead of two
(this is precisely what happens for a ferromagnet).

\medskip
  One way of understanding this is as follows. A classical complex
  field only represents one \ac{dof} since \(\varphi\) and
  \(\varphi^*\) are canonically conjugate to each other. Taking the
  large-\(\mu\) limit in the non-Abelian sector can thus be interpreted in
  two equivalent ways. Either we say that in a relativistic system we
  disregard the effect of the massive mode (mass \(\order{\mu}\)) or
  then we say that we go to a non-relativistic configuration. In both
  cases we must end up with only one Goldstone field, with dispersion
  \(\omega^2 \propto p^4\).

\medskip
  In the following section we show how the presence of the \ac{vev}
  \(v\) changes this result in the Abelian sector, where we can no
  longer take the limit to a non-relativistic field theory. We find a
  massive mode and a relativistic Goldstone mode, albeit propagating
  with speed \(c < 1\).

\subsection{Canonical Quantization of the Abelian sector}

In this section we move on to the Abelian sector. We diagonalize the quadratic Hamiltonian resulting from expanding around $v$ through a generalized Bogoliubov--Valatin 
transformation (see \emph{e.g.}~\cite{Xiao:1198089}). With that we prove the existence of the previously discussed  gapped modes.

In the $O(2)$ sector, the Lagrangian density quadratic in the fluctuating fields $(\phi_{2k-1}\,,\,\phi_{2k})$ reads
\begin{multline}
  \mathcal L^{(2)} = \frac{1}{2} \left( \partial^\mu \phi_{2k-1} \partial_\mu \phi_{2k-1} +  \partial^\mu \phi_{2k} \partial_\mu \phi_{2k}  \right) \\
  -\mu^2 \frac{2c^2}{1-c^2} \phi_{2k-1}^2
  +\mu \left( \phi_{2k}\dot\phi_{2k-1} - \phi_{2k-1} \dot\phi_{2k} \right)
  \,.
\end{multline}
The  \ac{eom} for the two fields are coupled and admit the solutions
\begin{equation}
  \phi_i(t,x) = \int \frac{\dd[d]{p}}{(2\pi)^d } \phi_i(t,p) e^{-ipx}
\end{equation}
with
\begin{align}
\phi_{2k-1}(t,p) &= \frac{\alpha a_{k}(p)}{\sqrt{p^2+\omega_+(p)^2}} e^{i\omega_- t}+ \frac{\beta b_{k}(p)}{\sqrt{p^2+\omega_-(p)^2}} e^{i\omega_+ t},\\
 \phi_{2k}(t,p) &= -\frac{i}{\sqrt{2}p} \left( \alpha a_{k}(p) e^{i\omega_- t}-\beta b_{k}(p) e^{i\omega_+ t}\right),
\end{align}
where $\alpha$ and $\beta$ are integration constants, $a_{k}(p)$ and $b_{k}(p)$ are generic functions of $p$, and 
\begin{equation}
  \omega_{\pm} \equiv \omega_{\pm}(p) =  %
 \sqrt{p^2 + \frac{2 \mu}{1-c^2}\pqty{\mu \pm \sqrt{\pqty{1-c^2}^2 p^2 + \mu^2}}} 
\end{equation}
gives the dispersion relation of the two modes in the Abelian sector.
The conjugate momenta to $\phi_{2k-1}\,,\,\phi_{2k}$ are 
\begin{align}
\pi_{2k-1} &= \dot \phi_{2k-1} + \mu \phi_{2k} , &
\pi_{2k} &= \dot \phi_{2k} - \mu \phi_{2k-1}
\,,
\end{align}
so that the corresponding Hamiltonian density is
\begin{multline}
  \mathcal H^{(2)} = \frac{1}{2} \Big[\pi_{2k-1}^2+\pi_{2k}^2+(\nabla\phi_{2k-1})^2+(\nabla\phi_{2k})^2
    +\mu^2\left(\frac{1+3c^2}{1-c^2}\phi_{2k-1}^2+\phi_{2k}^2\right)\\
-\mu(\pi_{2k-1}\phi_{2k}-\pi_{2k}\phi_{2k-1}) \Big]
\,.
\end{multline}

To quantize this Hamiltonian we proceed, as usual, by  promoting  $\phi_i$ and $\pi_i$  to operators satisfying the canonical equal-time commutation relations
\begin{align}
\label{Abelian:EqualTimeCommutator}
\comm{\phi_i(t,x)}{\pi_j(t,y)} &= i \delta_{ij}\,\delta^{(d)}(x-y) &
i,j = {2k-1,k} \,,
\end{align}
which is to say, we promote  $a_k(p)$ and \(b_k(p)\)  to Heisenberg operators:
\begin{align}
  \comm{a_k(p)}{a_k^\dagger(p')} &= \delta(p-p')\,, &   \comm{b_k(p)}{b_k^\dagger(p')} &= \delta(p-p')\,, &
  \comm{a_k(p)}{b_k^\dagger(p')} &= 0 \,.
\end{align}
We can now write the field operators starting from the classical
solution, imposing that the fields are real and canonically commute,
\begin{align}
  &\begin{multlined}[][.85\textwidth]
 \phi_{2k-1}(p) =\frac{\delta}{\sqrt 2} \bigg[
-\sqrt{ \frac{p^2-\omega_-^2}{\omega_-} }\left(a_{k}(p)+a_{k}^\dagger(-p)\right) \\+ 
\sqrt{\frac{\omega_+^2 - p^2}{\omega_+} } \left(b_{k}(p)+b_{k}^\dagger(-p)\right) \bigg] \,,    
  \end{multlined}
\\
  &
    \begin{multlined}[][.85\textwidth]
      \phi_{2k}(p) = i\frac{\delta}{\sqrt{2}p} \bigg[ \sqrt{\omega_-
          \left(\omega_+^2 - p^2 \right)} \left(a_{k}(p) -
          a_{k}^\dagger(-p)\right) \\
        + \sqrt{\omega_+ \left(p^2 - \omega_-^2 \right)} (b_{k}(p) -
        b_{k}^\dagger(-p))\bigg] \,.
    \end{multlined}
\end{align}
We still have an overall normalization constant $\delta$, which will be fixed by diagonalizing the Hamiltonian in the oscillators.
The commutation relation \eqref{Abelian:EqualTimeCommutator}
fixes the form of the canonically conjugate operators $\pi_i(k)$\,,
\begin{align}
 & 
   \begin{multlined}[][.85\textwidth]
     \pi_{2k-1}(p) =\frac{i}{ 2 \delta\sqrt{2}p^2 } \bigg[ \frac{\omega_+^2 + p^2 }{\omega_+^2 - \omega_-^2} \sqrt{ \omega_-\left(p^2 - \omega_-^2\right)}  \left(a_{k}(p) - a_{k}^\dagger(-p)  \right) \\
     -\frac{p^2 +\omega_-^2}{\omega_+^2 - \omega_-^2} \sqrt{\omega_+ \left( \omega_+^2 -
         p^2 \right) } \left(b_{k}(p) - b_{k}^\dagger(p) \right)
     \bigg] ,
   \end{multlined} \\
  &
    \begin{multlined}[][.87\textwidth]
      \pi_{2k}(p) = \frac{1}{2 \delta\sqrt{2} p} \bigg[ \frac{ p^2 +
        \omega_-^2}{\omega_+^2 - \omega_-^2} \sqrt{\frac{\omega_+^2 - p^2}{\omega_-} }
      \left(a_{k}(p) + a_{k}^\dagger(-p) \right)      \\
      +\frac{\omega_+^2 + p^2}{\omega_+^2 - \omega_-^2} \sqrt{\frac{p^2-
          \omega_-^2}{\omega_+} } \left(b_{k}(p) +
        b_{k}^\dagger(p)\right) \bigg] .
    \end{multlined}
\end{align}
The linear basis change in oscillator space is solely expressed  in terms of $\omega_\pm$ and $p$, but depends only implicitly  on $\mu$ and $c$. Consequently, the form of the transformation matrix does not change for generic potential $V$.

Eventually, substituting our ansatz in the Hamiltonian, we find that for
\begin{equation}
  \delta = {\pqty{\omega_+^2-\omega_-^2}^{-1/2}}  \,,
\end{equation}
the Hamiltonian is diagonal in the oscillators:
\begin{equation}
\label{O2:QuadraticHamiltonian_Diagonal}
  \begin{aligned}
    \mathcal{H}^{(2)} &= \omega_-(p) a_{k}^\dagger(p) a_{k}(p) + \omega_+(p)
    b_{k}^\dagger(p) b_{k} (p)
    \\
    &= c p \, a_{k}^\dagger(p) a_{k}(p) + \frac{2\mu}{\sqrt{1-c^2} }\,
    b_{k}^\dagger(p) b_{k}(p) + \order{\frac{1}{\mu}}
    \,.
  \end{aligned}
\end{equation}
This shows that \(a_k\) corresponds to a Goldstone with dispersion
\(\omega_-(p)\), \emph{i.e.} a phonon with velocity \(c\)\,, while \(b_k\)
represents a massive mode with dispersion \(\omega_+\).

Going back to the fields \(\phi_i\), their large-$\mu$ expansion is given by
\begin{align}\label{eq:phi2k1}
  \phi_{2k-1}(p) &\sim %
                   \frac{(1-c^2)^{1/4}}{2\sqrt{\mu}} \left( b_{k}(p) + b_{k}^\dagger(-p) \right)
                   - \frac{1 -c^2}{2c } \frac{p}{\mu} \sqrt{\frac{c}{2p} } \left( a_{k}(p) + a_{k}^\dagger(-p) \right),
   \\
  \phi_{2k}(p) &\sim %
i\sqrt{\frac{c}{2p} }   \left( a_{k}(p) - a_{k}^\dagger(-p) \right) + i \frac{(1-c^2)^{3/4}}{2\sqrt{\mu}} %
 \left( b_{k}(p) - b_{k}^\dagger(-p) \right)\label{eq:phi2k}
 \,.
\end{align}
As expected, we see that at lowest order, \(\phi_{2k}\) behaves like a
Goldstone, while \(\phi_{2k-1}\) behaves like a massive field.
The canonical commutation relations are satisfied to each order in $\mu$, proving the consistency of the expansion.

The Abelian sector behaves more like the antiferromagnetic case, where the Berry’s phase term merely changes the spin wave velocity and does not affect the spectrum qualitatively. 

\medskip
Up to this point, our result is independent of the choice of
parametrization of the fluctuations that we discussed in the previous
section. This changes once we consider the interactions. In the choice
\(\phi = e^{i \mu t + i \varphi_2/v} (v + \varphi_1)\), the potential only
depends on \(\varphi_1\) and everything is fine, because \(\varphi_1\)
starts at order \(1/\sqrt{\mu}\) and the Goldstone appears multiplied
by \(k\) at order \(1/\mu\). In the other choice, there is no control:
the potential depends on \(\varphi_2\), which is order \(1\) and as a
result we get infinite terms of the same order in the Dyson expansion.

\section{Suppression of the interactions}\label{sec:supp}

In this section, we want to show that all interaction terms are suppressed by $\mu$ (which is large at large charge) for a general potential of the form $V\propto \abs{\varphi}^m$ with $m \ge 2$ in $d$ space-time dimensions. In order for a condensate to exist, we necessarily need to work  in $d>2$. For convenience, we use $\mu$ and $v$, which are both functions of $\bar\rho$, which is by construction the only dominant scale. 

\medskip
Up to this point, we have assumed that the quadratic part of the Hamiltonian is the most important and that the rest can be treated as small. After having diagonalized $\mathcal{H}^{(2)}$,  we can come back to this assumption and verify it using the expansion of the fields in terms of Goldstones and massive operators. At leading order in $\mu$ and therefore also in $\bar\rho$, the field $\phi_{2k}$ corresponds to the relativistic Goldstone boson. Since due to the $O(2n)$ invariance, $V(\phi)$ does not depend on $\phi_{2k}$, there are only two higher order terms that involve the relativistic Goldstone. They are of the type 
\begin{align}
v\phi_{2k-1}\frac{\phi_{2k}^2}{v^2} && \text{and} && \phi_{2k-1}^2\frac{\phi_{2k}^2}{v^2} \,.
\end{align}
Expanding in oscillators (see Eq.~\eqref{eq:phi2k1} and \eqref{eq:phi2k}), we see that the first term goes like $(v\sqrt\mu)^{-1}$ and the second like $(v^2\sqrt\mu)^{-1}$. They both correct the propagator of the Goldstone by a term $(v^2\mu)^{-1}\ll 1$.

\medskip

In order to be able to compare the interaction term to the quadratic part, we expand the potential 
\begin{equation}
V(\phi) = V(v^2)+\mu^2 \lambda^{i_1 i_2} \varphi_{i_1} \varphi_{i_2} + \mu^2 \frac{\lambda^{i_1 i_2 i_3}}{v}\varphi_{i_1} \varphi_{i_2} \varphi_{i_3} + \dots + \mu^2 \frac{\lambda^{i_1\dots i_m}}{v^{m-2}}\varphi_{i_1} \dots \varphi_{i_m} \,,
\end{equation}
where the \(\lambda\) are dimensionless constants and of order \(\order{1}\).
To first approximation, when expressed in terms of Heisenberg
operators Eq.~\eqref{eq:varphi}, $\varphi_i$ is of order $\order{\mu^{-1/2}}$ so the
interaction terms among \(m\) fields \(\set{\phi_i, i = 1, \dots, 2k -
  1}\) become
\begin{equation}
  \frac{\mu^2 \lambda^{i_1\dots i_m}}{v^{m-2}\mu^{m/2}} = \frac{\lambda^{i_1\dots i_m}}{v^{m-2}\mu^{m/2-2}} \,.
\end{equation}
$v$ has the dimensions of a field, \([v]= d/2-1\), so overall we have
\begin{equation}\label{eq:dims}
  \frac{\lambda^{i_1\dots i_m}}{\mu^{-d+m/2(d-1)}} = \frac{\lambda^{i_1\dots i_m}}{\mu^{\Omega_m}} \,.
\end{equation}
Using $[\bar\rho]=d-1$, we find that Eq.~\eqref{eq:dims} in terms of $\bar\rho$ is given by
\begin{equation}
  \frac{\lambda^{i_1\dots i_m}}{\bar\rho^{(m/2-d/(d-1))}} \,.
\end{equation}
For $m\geq 4$, 
\begin{equation}
  \Omega_m = \tfrac{m}{2} \pqty{d-1} - d > 0
\end{equation}
and the interactions are suppressed.
The only term that is possibly not suppressed arises for $d=3$, $m=3$, since $\Omega_3 = \tfrac{1}{2}(d-3)$.
Given our choice of the \ac{vev}, $\expval{\varphi} = (0, \dots, v)$, the cubic term can either be of the form
\begin{align}
\phi^3_{2k-1} &&
\text{or} && \phi_{2k-1}\varphi_i^2 \,, \hspace{2em} i=1,\dots,k-1 \,.
\end{align}
In either case, they lead to corrections to the mass of $\phi_{2k-1}$, which is of order $\order{\mu}$.

\section{Calculating the anomalous dimension}\label{sec:confdim}
 
The fixed-charge ground states around which we are expanding depend explicitly on time and violate Lorentz invariance. Note, that we have assumed so far that the theory lived on $\mathbb{R}_t \times \mathbb{R}^{d-1}$\,, but of course we could have chosen a more general background of the type $\mathbb{R}_t \times \mathcal{M}^{d-1}$ with only minor changes, $\mathcal{M}^{d-1}$ a ${d-1}$-dimensional manifold. In the particular case of a conformal theory, the choice of the $(d-1)$-sphere $\mathcal{M}^{d-1}=S^{d-1}$ leads to an interesting application of our general construction.

We are assuming that we still have the Goldstone structure, which is strictly true only in the infinite-volume limit. As long as we assume that our effective action is valid for energies $\Lambda \gg 1/R$, this assumption is however justified.

In order to calculate the conformal dimension, we need to first perform an analytic continuation, $t \to i\tau$. $\mathbb{R}\times S^{d-1}(r_0)$ is conformally flat, with metric
\begin{equation}
  \dd s^2 = \dd{\tau}^2 + r_0^2 \dd{\Omega^2_{d-1 }} = \frac{r_0^2}{r^2} \pqty{\dd{ r^2} + r^2 \dd {\Omega^2_{d-1}}} \,,
\end{equation}
where $r=r_0\,e^{\tau/r_0}$.
Our initial time coordinate has now become the radius $r$ and the
Hamiltonian is identified with the dilatation operator. In other
words, a state with fixed charge and energy $E$ on $\mathbb{R}_t\times S^{d-1}$ is mapped to an operator on $\mathbb{R}^{d}$ with conformal dimension 
\begin{equation}
D= r_0\, E \,.
\end{equation}

In the following, we shall consider the case $d=3$ and repeat our construction for the $O(2n)$ model, fixing the values of all $n$ charges,
\begin{equation}
\overline Q_i = \int_{S^2} r_0^2  \bar\rho_i \dd{ \Omega} = 4\pi r_0^2\, \bar \rho_i
\,.
\end{equation}
In this context, the action \eqref{eq:O2n_LagrangianDensity} for $2n$ real scalar fields $\phi_a$ on $\mathbb{R}_t\times S^{d-1}$, conformally coupled to the metric, is given by
\begin{equation}
  \label{eq:O(n)-on-the-sphere}
S=\tfrac{1}{2} \int \dd t \, r_0^2 \, \dd \Omega \left[ \del_\mu \phi_a g^{\mu\nu} \del_\nu \phi_a-  %
V(\phi_a\phi_a)
\right] ,
\end{equation}
where the potential becomes now
\begin{equation}
V(\phi_a \phi_a) = \sum_{i=1}^{2n} \left(\xi R \phi_a^2 + \frac{\lambda}{3}\,\phi_a^6\right),
\end{equation}
with $\xi=\tfrac{1}{4} \pqty{d-2}/\pqty{d-1} = {1}/{8}$ and the Ricci scalar $R={2}/{r_0^2}$\,.
For this configuration, we immediately find  that in the ground state,
\begin{align}
\mu &= \left( \frac{\xi R+\sqrt{\xi^2R^2+4\lambda\bar\rho^2}}{2}\right)^{1/2}  = \lambda^{1/4}\bar\rho^{1/2}+\order{1/\bar\rho}\,,\\
  v &= \left( \frac{-\xi R+\sqrt{\xi^2R^2+4\lambda\bar\rho^2}}{2\lambda}\right)^{1/4} \sim \lambda^{-1/8}\,\bar\rho^{1/4}+\order{\bar\rho^{-3/2}}\,, \\
  c^2 &= \frac{1}{2} \pqty{1 - \frac{R \xi}{\sqrt{\xi^2 R^2 + 4 \lambda \bar \rho^2}}} = \frac{1}{2} + \order{\bar \rho^{-1}}
  \,,
\end{align}
while the energy of the configuration is
\begin{equation}
  \begin{aligned}
    E_0 &= 4\pi r_0^2\left[  \frac{1}{3\sqrt{2\lambda}}\left(2\xi R+\sqrt{\xi^2R^2+4\lambda\bar \rho^2}\right) \left(\sqrt{\xi^2R^2+4\lambda\bar \rho^2}-\xi R \right)^{1/2} \right]\\
& = 4\pi r_0^2 \left( \tfrac{2}{3}\lambda^{1/4}\bar\rho^{3/2} + \frac{\xi R}{2 \lambda^{1/4}} \sqrt{\bar\rho} + \order{\bar\rho^{-1/2}}\right).
\end{aligned}
\end{equation}

The analysis of the fluctuation proceeds in parallel to the one in flat space explained in the previous sections. The only difference is that now, the fields are expanded into spherical harmonics $Y_{l,m}$ according to
\begin{align}
\phi_i(t,\Omega) &= \sum_{l,m}\frac{1}{\sqrt{2\omega}} \left[ e^{i\pi |m|/2} e^{-i\omega t} Y_{l,m}(\Omega) a_i(l,m)+ e^{-i\pi |m|/2} e^{i\omega t} Y^*_{l,m}(\Omega) a_i^\dagger(l,m)\right],
\end{align}
where both
$l,m \in \mathbb Z$ and $l \geq 0$ while $m=-l,...,0,...,l$\,.
The $a_i(l,m)$ for $i=1,...,2n$ satisfy the standard commutation relations
\begin{align}
 \comm{a_i(l,m)}{a_j^\dagger(l',m')} = \delta_{l l'} \delta_{m m'} \, \delta_{ij}
 \,,
\end{align}
as before. Using this expansion for the real fields, it turns out that
the expression for the Hamiltonian on $S^2$ is formally the same as
the one in flat space after the following substitutions are performed:
\begin{align}
a(p) &~\mapsto~ a(l,m)\\
p^2 &~\mapsto~ \frac{l(l+1)}{r_0^2}
\,,
\end{align}
This means that we still have a relativistic Goldstone with dispersion relation 
\begin{equation}
  \begin{aligned}
    \omega_{-} &= \sqrt{\frac{l(l+1)}{r_0^2} + \frac{2 \mu}{1-c^2}\pqty{\mu -\sqrt{\pqty{1-c^2}^2 \frac{l(l+1)}{r_0^2} + \mu^2}}}  \\
    & =\frac{1}{\sqrt 2 r_0} \sqrt{l (l+1)} + \order{\bar \rho^{-1}} \,.
  \end{aligned}
\end{equation}
Its contribution to the energy is proportional to the Casimir energy
on $S^2$~\cite{Monin:2016bwf}:\footnote{This corrects a
  mistake in the regularization made in~\cite{Hellerman:2015nra}.}
\begin{equation}
  E_{G}=\frac{1}{2\sqrt{2}r_0} \left( -\frac{1}{4} - 0.015  \right) .
\end{equation}
The non-relativistic Goldstones do not contribute because in the large-$\mu$ limit, they are classical fields and the Hamiltonian annihilates the vacuum.

All other contributions are suppressed by $\overline Q^{-1/2}$, as shown in the previous section. Adding the contribution of the condensate to the one of the Goldstone, we can
evaluate the dominant terms in the large-charge expansion of the conformal dimension:
\begin{equation}
  \label{eq:conformal-dimension-formula}
  \begin{aligned}
    D(Q) &= r_0 (E_0 + E_G) = \frac{\lambda^{1/4}}{3\sqrt{\pi}} \overline Q^{3/2}+\frac{\sqrt{\pi}}{4\lambda^{1/4}} \overline Q^{1/2} - 0.093 + \order{\overline Q^{-1/2}}\\
&= \alpha_{3/2} \overline Q^{3/2} + \frac{1}{12 \alpha_{3/2}} \overline Q^{1/2} - 0.093 + \order{\overline Q^{-1/2}} \,.
\end{aligned}
\end{equation}
We find a form that is universal for all $O(n)$ models, depending on a
single parameter $\alpha_{3/2}$ that can be determined \emph{e.g.}
from \ac{MC} computations.

\begin{figure}
  \centering
  \includegraphics[width=.8\textwidth]{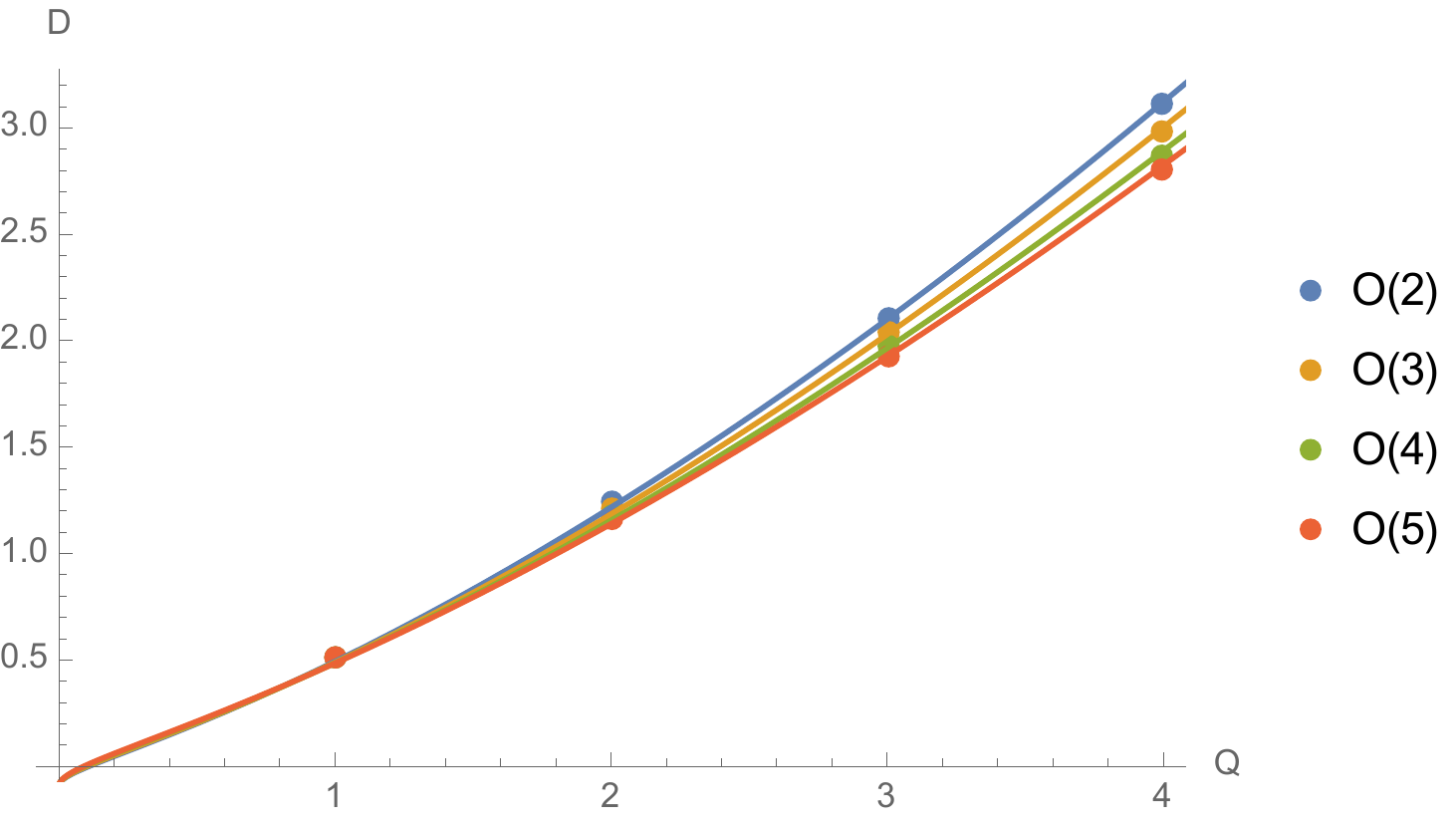}
  \caption{$D(Q)$ from \ac{MC} simulations in $O(n)$,
    $n=2,\dots,5$~\cite{Hasenbusch}. The continuous lines are
    one-parameter fits for the formula in
    Eq.~\eqref{eq:conformal-dimension-formula} with  $\alpha_{3/2}=0.34,\,0.32,\,0.30,\,0.29$.}
  \label{fig:MC-data}
\end{figure}

The plot in Figure~\ref{fig:MC-data} shows the values of the conformal
dimension $D(Q)$ stemming from \ac{MC} simulations in $O(n)$,
$n=2,3,4,5$~\cite{Hasenbusch}. The continuous lines are one-parameter
fits for $\alpha_{3/2}$ which is quite good even though the values of
$Q$ are small.
Using the values of \(\alpha_{3/2}\) coming from the fit, we can compute \(\lambda\) and
verify our assumptions. For \(O(n)\), \(2 \le n \le 5\) we
find that coupling \(2 \gtrsim \lambda/3 \gtrsim 3.5\). As expected, the coupling
is of order \(\order{1}\); in other words, we are in a regime where standard perturbation theory
would be useless.

The action in Eq.~\eqref{eq:O(n)-on-the-sphere} is not the most
  general one compatible with the symmetries of the problem. In fact,
  in terms of the fields \(\phi_{2k-1}\) and \(\phi_{2k}\), we could
  have started with \(1/2 \pqty{\del_\mu \phi_{2k-1}}^2 + b/2
  \phi_{2k-1}^2 \pqty{\del_\mu \phi_{2k}}^2\), where \(b\) is an arbitrary
  parameter. In the spirit of~\cite{Hellerman:2015nra}, this gives the
  effective Wilsonian action describing the conformal fixed point of
  the \(O(n)\) model in the limit of large charge\footnote{We would
    like to thank Simeon Hellerman for discussions about this point.}. Repeating the
  computations we find that the coefficients \(\alpha_{3/2}\) and
  \(\alpha_{1/2}\) are independent and their product is \(\alpha_{3/2}
  \alpha_{1/2} = 1/(12 b)\). Interestingly enough, though,
  fitting the values for the conformal dimensions stemming from
  \ac{MC} simulations, shows that \emph{empirically} the product
  \(\alpha_{3/2} \alpha_{1/2}\) is compatible with the value \(b=1\)
  that we used above. We intend to revisit the question of this apparent
  coincidence in future work.
\section*{Acknowledgments}
 The authors would like to thank Antonio Amariti, Matthias Blau,
 Simeon Hellerman, Mikko Laine, Slava Rytchkov and Uwe--Jens Wiese for
 enlightening discussions and
 comments, and {Sean Hartnoll for pointing out some improvements}.  
 We would also like to thank the anonymous referee for suggesting improvements to the introduction.
 D.O. and S.R. gratefully acknowledge support from the Simons Center for Geometry and Physics, Stony Brook University at which some of the research for this paper was performed.
The work of S.R. and O.L. is supported by the Swiss National Science Foundation (\textsc{snf}) under grant number \textsc{pp}00\textsc{p}2\_157571/1.

\printbibliography

\end{document}